\documentclass[useAMS,usenatbib,usegraphicx]{mn2e}

\title[Dating massive early-type galaxies at $z\sim$1.5]
{Dating the stellar population in massive early-type galaxies 
at $z\sim$1.5}
\author[M. Longhetti et al.]{M. Longhetti$^{1}$\thanks{E-mail:
marcella@brera.mi.astro.it}, P. Saracco$^{1}$, P. Severgnini$^{1}$,
R. Della Ceca$^{1}$, V. Braito$^{1}$, 
\newauthor F. Mannucci$^{2}$, R. Bender$^{3,4}$,
N. Drory$^{5}$, G. Feulner$^{3}$, U. Hopp$^{3}$ \\
$^{1}$INAF - Osservatorio Astronomico di Brera, Via Brera 28, 20121 Milano\\
$^{2}$IRA-CNR, Firenze, Italy\\
$^{3}$Universit\"ats-Sternwarte M\"unchen, Scheinerstr.~1, D-81679
M\"unchen, Germany\\
$^{4}$Max-Planck-Institut f\"ur extraterrestrische Physik,
Giessenbachstra\ss e, D-85748 Garching, Germany\\
$^{5}$University of Texas at Austin, Austin, Texas 78712
}
\begin{document}

\date{Accepted 2005. Received 2005; in original form 2005}
\pagerange{\pageref{firstpage}--\pageref{lastpage}} \pubyear{2005}
\maketitle
\label{firstpage}
\begin{abstract}
We present the analysis of 10 massive early-type galaxies at $z\sim1.5$.
They have been identified by means of a near-IR low resolution
spectroscopic follow-up of a complete sample of 36 bright (K' $<$ 18.5) 
Extremely Red Objects (EROs, R-K'$>$ 5) selected from the Munich Near-IR Cluster
Survey  (MUNICS; Drory et al. 2001).
The low resolution near-IR spectra constrain their redshift
at $1.2<z<1.7$, implying absolute magnitudes M$_{K'}<-26.0$ and
stellar masses well in excess of 10$^{11}$ M$_\odot$.
Under the hypothesis of pure passive evolution
from $z\sim1.5$ to $z=0$, in the local universe they would 
have luminosities L$_K\ge2.5$L$^*$.
Thus, they are the high-z counterparts of the local old massive 
(10$^{11}-10^{12}$ M$_\odot$) early-type galaxies populating the bright end of 
the local luminosity function of galaxies.
The comparison of their spectro-photometric properties
with a grid of synthetic models suggests that the stellar populations
in more than half of the sample are about $\sim$3-5 Gyr old and 1-2 Gyr old
in the remaining part.
These ages imply formation redshift $z_{f} > 2$ for all the galaxies
and $z_{f} \geq 4$ for the oldest ones.
The comparison of the 4000\AA\  break and of the overall spectral shape of
the average spectrum of the 10 galaxies at $z\sim1.5$ with those of their local counterparts
confirms that field massive early-type galaxies
formed the bulk of their stellar mass at $2<z<4$, most likely over
a short ($<$ 1 Gyr) star formation time scale, consistently with the
results derived from the analysis of their individual spectro-photometric properties.

\end{abstract}

\section{Introduction}
The epoch of formation of high mass ($\mathcal{M}_{star}>10^{11}$ M$_\odot$)
early-type galaxies represents a key issue to understand the whole picture
of galaxy formation.

\noindent
The formation and the evolution of galaxies are mainly the result of two physical
processes: the mass assembly and the conversion of gas into stars.
By now, the mass assembly on large scale is well described by
the hierarchical models of galaxy formation  in which
structures are formed
by means of subsequent merging of dark matter haloes. Observational
confirmation of this theoretical framework and precise constraints
on its relevant cosmological parameters came recently from the analysis
of the WMAP ({\it Wilkinson Microwave Anisotropy Probe}, Bennett et al.
2003) data (e.g., Spergel et al. 2003).
On the other hand, the link between the dark matter distribution at
any epoch and the properties of the corresponding baryonic luminous
matter is not straightforward, and the semianalytic models aimed at
building this link are still not exhaustive.
One of the difficulties of the
hierarchical models of galaxy formation is related to
the properties and the evolution  of the population of
massive galaxies.  This class of objects is locally composed mainly
of early-type galaxies, i.e. ellipticals and spheroids.
Indeed, in the hierarchical models spheroids are
assembled by means of subsequent mergers of smaller disk galaxies
occurring at $z < 2$  (Kauffmann \& Charlot 1998;
Baugh et al. 2003). This implies evolution
with $z$ of the co-moving space density of massive spheroids:
the higher the redshift, the lower the mass
they can have already assembled. In particular, the most massive
spheroids ($M_{star} \sim 10^{11}-10^{12} M_{\odot}$), populating
the bright end of the local luminosity function, reach
their final mass at $z \sim 1$ depending on the chosen values
of some parameters in the models.
On the other hand, there is evidence that fully assembled
massive galaxies exist at $z > 1$
(Saracco et al. 2005; Cimatti et al. 2004; McCarthy et al. 2004;
van Dokkum et al. 2004; van Dokkum \& Stanford 2001) and the properties
of their stellar content suggest they are evolved
structures, i.e. they are early-type galaxies with old
stars which can only passively evolve
in the local population of massive spheroids.
The observations of massive galaxies at $z>1$ can be more easily
taken into account in the monolithic scenario of galaxy
formation, that forms even the most massive
ellipticals at high
redshift ($z > 2-3$) in a single episode of mass collapse.
In this framework, 
at the origin of massive spheroids
all the gas is burned into stars during the mass
collapse, and the luminosity evolution of the
galaxies follows a
pure passive aging (e.g. Tinsley 1977;
 Bruzual \& Kron 1980).
Both the hierarchical and the monolithic scenarios are in agreement 
with the observational evidence
that the bulk of the stellar populations in local ellipticals and
spheroids is relatively old (i.e., consistent with a formation redshift
$z_f > 3$: e.g.,  Thomas et al. 2005, Renzini \& Cimatti 1999)
while differences in some foreseen properties come out at redshift $z > 1$.
Therefore, the age and the properties of the stellar populations 
in field massive early-types
at $z>1$ can be a key test for the galaxy 
formation models.
Indeed, by determining the mean age of the stellar content of massive 
field early-types 
it is possible to constrain their star formation history (SFH) and the epoch of
their formation by tracing forward their evolution (over an interval
of about 7-8 Gyr)  to match the properties of the local massive 
early-type galaxies.

It is worth to note that $\mathcal{M}_{star}>10^{11}$ M$_\odot$ 
early-types populate the very bright end (L$>$L$^*$) of the local 
luminosity function of galaxies. 
Thus, $z>1$ early-type galaxies with these high stellar masses have necessarily 
already completed the accretion of their stellar mass  at that redshift and, 
therefore, are the high-z counterpart of the local old ones.

In this paper we report the analysis of a sample of 10 high-mass 
($10^{11}<\mathcal{M}_{star}<10^{12}$ M$_\odot$) 
field galaxies spectroscopically classified as early-type 
galaxies at $1.2 <z< 1.7$. 
They result from an on-going near-IR spectroscopic follow-up  
(Saracco et al. 2003, 2005) of a complete
sample of 36 bright (K' $<$ 18.5) Extremely Red Objects (EROs, R-K'$>$ 5)
selected over two fields ($\sim$320 arcmin$^2$) of the Munich Near-IR Cluster 
Survey  (MUNICS; Drory et al. 2001).
Optical  (B, V, R, I) and near IR (J and K') photometry is 
also available for the sample.
Furthermore, the two fields have been targets of
two {\it XMM-Newton} pointings. One of
them has been already analyzed and it has allowed the
detection of 6  X-ray emitting EROs, five of which  
have properties matching those expected for X-ray obscured type-2 QSO
(Severgnini et al. 2005).

We present the analysis of the  spectro-photometric data
of the 10 early-type galaxies identified so far
in \S2, where we derive their redshift and luminosities.
In \S3 we derive a robust estimate of their stellar mass content
and of the age of the bulk of their stars through the comparison with synthetic models.
Finally, we discuss the derived properties in terms of
galaxy formation and evolution scenarios in \S4 and 
we summarize the  results in \S5. 
The comoving density of high-mass early-type galaxies is presented
and discussed in another paper (Saracco et al. 2005).  
The morphological analysis of the 10 galaxies, based on 
H band NICMOS-HST data (Cycle 14) will be presented in a further paper.
Throughout this paper we assume H$_0$=70 km s$^{-1}$
Mpc$^{-1}$, $\Omega_M=0.3$ and $\Omega_{\Lambda}=0.7$.
All magnitudes are given in the Vega system.

\section[]{The near-IR spectra}

Near-IR spectroscopic data have been collected in October 2002
and November 2003 at the Italian 3.6 m telescope TNG 
(Telescopio Nazionale Galileo, LaPalma,
Canary Islands) using the AMICI prism mounted at the NICS spectrometer,
with typical exposure of 3-4 hours on source. 
The prism provides the spectrum from 0.85 $\mu$m to 2.4 $\mu$m in
a single shot with a nearly constant resolution of 
$\lambda/\Delta\lambda\simeq35$ (1.5'' width slit).
The resulting 
spectral dispersion is $\sim$30\AA\ (100\AA) per pixel, while the
Full-Width at Half-Maximum (FWHM) is $\sim 280$\AA\ ($\sim 570$\AA)
at 10000\AA\ (20000\AA).
This extremely low resolution is well suited to describe the spectral shape of 
the sources and to detect strong continuum features such as the 4000\AA\ break in
old stellar systems at $z>1.1-1.2$. On the other hand, it makes unfeasible the detection of
emission/absorption lines in sources as faint as the EROs presented here.

\begin{table*}
\caption{Photometry from the MUNICS catalog of the 10 early-type galaxies. 
All magnitudes are in the Vega system and measured within 5'' diameter 
aperture (Drory et al. 2001). Note: S7F5\_45\ has not been observed in the B-band.
}
\centerline{
\begin{tabular}{lcccccc}
\hline
\hline
  Object  & B&  V& R& I& J& K'\\
  \hline
S2F5\_109 & 24.2$\pm$0.3&23.5$\pm$0.2&21.8$\pm$0.1&20.1$\pm$0.1&18.2$\pm$0.1&16.6$\pm$0.1\\
S7F5\_254 & $>$25.0&$>$24.5&$>$24.0&23.1$\pm$0.7&19.8$\pm$0.1&17.8$\pm$0.1\\
S2F1\_357 & $>$25.0&$>$24.5&23.8$\pm$0.2&21.5$\pm$0.2&19.5$\pm$0.1&17.8$\pm$0.1\\
S2F1\_527 & $>$25.0&$>$24.5&$>$24.0&22.6$\pm$0.4&20.4$\pm$0.2&18.3$\pm$0.1\\
S2F1\_389 & 24.3$\pm$0.5&23.7$\pm$0.5&23.7$\pm$0.5&23.0$\pm$0.5&20.3$\pm$0.2&18.2$\pm$0.1\\
S2F1\_511 &  $>$25.0&$>$24.5&$>$24.0& 21.6$\pm$0.6&19.8$\pm$0.1&18.1$\pm$0.1\\
S2F1\_142 & $>$25.0&$>$24.5&23.8$\pm$0.3&21.5$\pm$0.2&19.6$\pm$0.1&17.8$\pm$0.1\\
S7F5\_45\  & - & 24.2$\pm$0.4&23.5$\pm$0.3&22.2$\pm$0.3&19.6$\pm$0.1&17.6$\pm$0.1\\
S2F1\_633 &  $>$25.0&$>$24.5&$>$24.0& 22.5$\pm$0.5&20.0$\pm$0.1&18.2$\pm$0.1\\
S2F1\_443 &  $>$25.0&$>$24.5&$>$24.0& 23.2$\pm$0.6&20.5$\pm$0.1&18.4$\pm$0.1\\
\hline
\hline
\end{tabular}
}
\end{table*}

\begin{figure}
 \centering
\includegraphics[width=8.8cm]{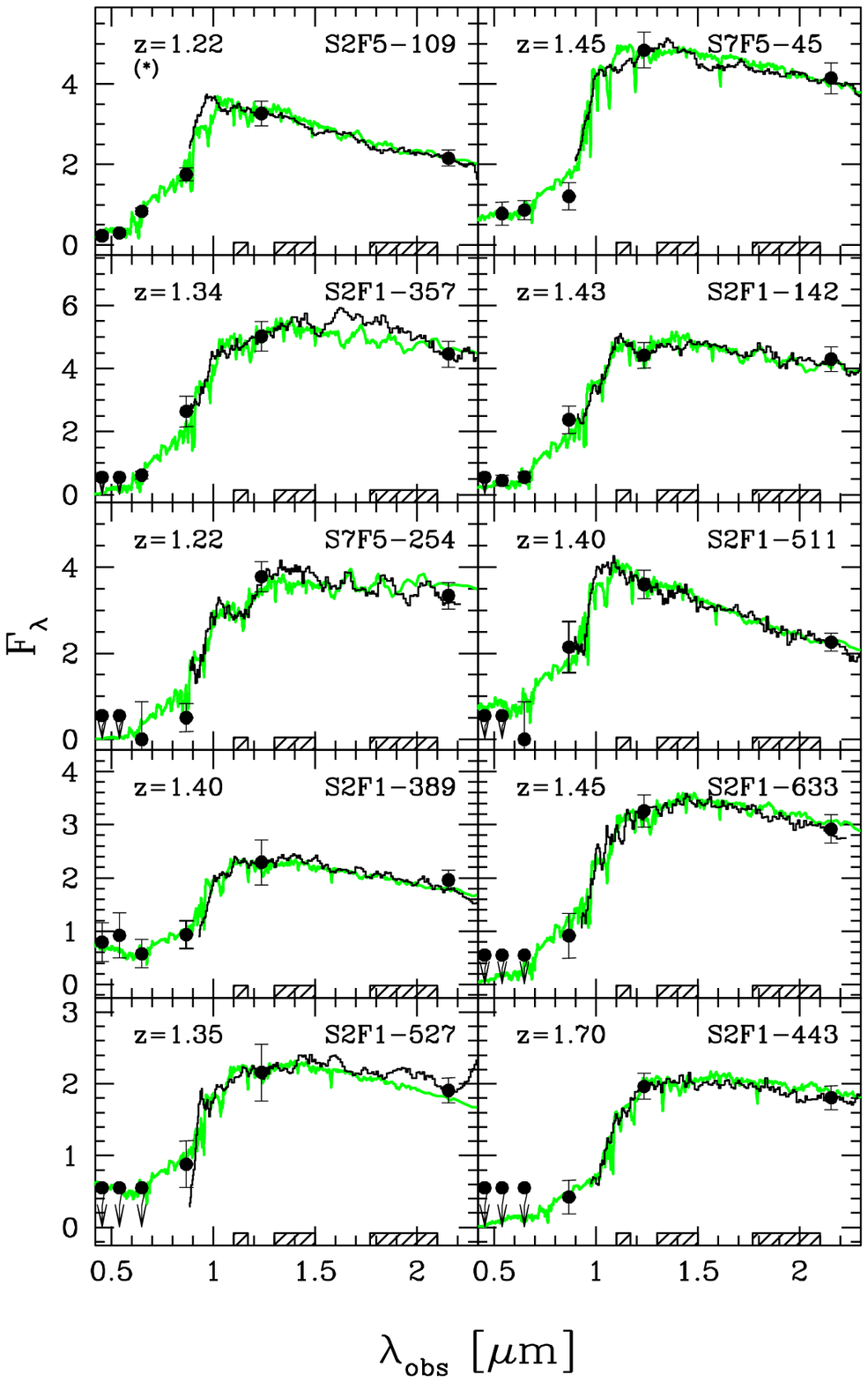}
 \caption{Near-IR spectra of the 10 early-type galaxies
(black line). Fluxes are  F$_{\lambda}$ [$10^{-18}$ erg s$^{-1}$ cm$^{-2}$ 
$\AA^{-1}$], while $(\star)$ indicates that the flux of S2F5\_109 is
reported in 5.0 10$^{-18}$ erg s$^{-1}$ cm$^{-2}$ \AA$^{-1}$ units.
The original NICS-Amici data have been 
smoothed to show the continuum superimposed on  synthetic
templates (green/grey line) normalized to their J-band flux. 
The shaded areas represent the spectral ranges characterized by atmospheric 
opacity larger than 80\%. 
The filled symbols are the photometric data in the B, V, R, I, J and K' bands from 
Drory et al. (2001; circles). 
In the upper left corner of each panel, the measured redshift
of each galaxy is reported ($\delta z$ = 0.05).}
\end{figure}

\noindent
In Figure 1 we show the smoothed spectra of the 10 early-type galaxies,
two of which (S7F5\_45 and S7F5\_254) previously analyzed by Saracco et al. 
(2003), while their broad band photometry is summarized in Table 1.
All the spectra drop very 
rapidly at wavelength $0.85-1.0\mu$m concurrent with the 4000\AA\ break.
The identification of the 4000\AA\ break in all the galaxies 
places them at $1.2\le z\le1.7$, while its steepness confirms their 
early-type spectral nature\footnote{In this context we refer to early-type
galaxies as galaxies whose stellar content has been mainly formed in a past 
star formation (SF) event and whose spectro-photometric properties 
can be accounted for only
by stellar populations older than 1.0 Gyr with small or absent dust
contribution.} 
In Saracco et al. (2005), the possibility that these EROs are dusty 
starburst galaxies has been already ruled out and details on the adopted
procedure are discussed.
Briefly, we searched for acceptable fits to the spectro-photometric data
of the EROs among a set of templates of starburst galaxies. 
The selected template library was made up by the six empirical starburst templates
(SB1-SB6) of Kinney et al. (1996) and by one starburst
model described by a constant star formation rate.
Extinction has been made varying in the range 0$<$E(B-V)$<2$.
For none of these 10 EROs we obtained an acceptable fit.

Considering the spectral width of the 4000\AA\ feature and the very low
resolution of our spectra, we measure the redshift of the 10
galaxies with an uncertainty of 0.05.
The resulting measured redshift are reported in  Figure 1 and summarized in Table 3.
We like to note that the highest redshift galaxy S2F1\_443 is one of the X-ray
emitting EROs detected on the S2F1 field (Severgnini et al. 2005).

The bright K'-band magnitudes (K'$<$ 18.4) of our early-type galaxies
together with their redshift $z > 1.2$ imply that their
stellar mass content is well in excess of $10^{11}$ M$_\odot$. 
Indeed, considering as reference the faintest magnitude
(K'=18.4) at the minimum redshift $z=1.2$, the corresponding rest-frame 
(k-corrected) luminosity
would be $\simeq 4\times10^{11}$ L$_\odot$
(M$_{\odot,K}=3.4$, Allen 1973). Assuming M$^*_K=-$24.3 mag for the
the local luminosity function (Kochanek et al. 2001),
it would be equivalent to $\simeq 3$ L$^{*}_{K}$.
Thus, even conservatively assuming  M/L$\simeq 0.5$
M$_\odot$/L$_\odot$ (i.e., half of the local value, Drory et al. 2004)
the resulting lowest stellar mass content of the present
sample of early-type galaxies would be $\ga 10^{11}$ M$_\odot$,
which is almost independent of any model assumption.

The simple analysis based on the near-IR spectra 
and the broad-band photometry of the 10 galaxies
has allowed us to constrain their redshift, to identify
their early-type nature and to derive their stellar mass in a model independent way.
Colours alone cannot provide the same strong constraints on their nature
and, most of all, on their redshift.
Indeed, in Figure 2 the R-K' colour of the 10 early-type galaxies
 is shown as a function of 
their J-K' colour. The selection criterion proposed by Pozzetti \& Mannucci (2000,
thin dashed lines) 
to discriminate early-type galaxies from starburst ones among the EROs is 
confirmed to be roughly predictive on their nature (see also
Cimatti et al. 2003). At the same time, at least three out of the ten
galaxies of our sample (namely, S2F1\_527, S2F1\_443 and S2F1\_389)
could not be correctly classified on the basis
of the Pozzetti \& Mannucci criterion, since their colours place them
on the border of the classification line. Actually, S2F1\_527 and S2F1\_443
are represented by lower limits in their (R-K') colour 
and are therefore located
in a lower region with respect to their real position in the diagram. 
This is not the case of S2F1\_389 that falls in the SB region
of the same diagram, even if quite close to the border line, and
whose early-type nature has been demonstrated by Saracco et al. (2005).
Indeed, its Spectral Energy Distribution (SED) from 0.4$\mu$m to 2.2$\mu$m
cannot be fitted by any starburst SED, both observed and synthetic one,
while it is very well reproduced by spectra of evolved stellar populations 
without any dust. 

Furthermore, in the diagram
we also show the colour-colour tracks expected for early-type galaxies.
Namely, we report the expected colours of two Simple Stellar Populations 
(SSPs) and one exponentially declining SFH with
time scale $\tau=0.6$ Gyr, based on Bruzual \& Charlot models (2003; BC03)
assuming the Salpeter Initial Mass Function (IMF) and solar metallicity.
It is evident the degeneracy of the expected colours with respect to the mean age of 
the stellar populations, the redshift and the SFH,
and consequently the low effectiveness of broad band colours in
constraining the properties of the stellar populations of galaxies.
On the contrary, the continuum shape over the rest-frame range 
3800-11000\AA\ provided by the near-IR spectra, even if at the very low 
resolution of our observations, makes it possible
to fix their redshift with reasonable accuracy. 
Furthermore, as it will be discussed in the next section,
when coupled with rest-frame UV data ($\lambda\le2800$ \AA) 
provided by the B and the V bands,  
the near-IR spectrum provides severe constraints on the age 
of the stellar population of our galaxies.

\begin{figure}
 \centering
 \includegraphics[width=8.5cm]{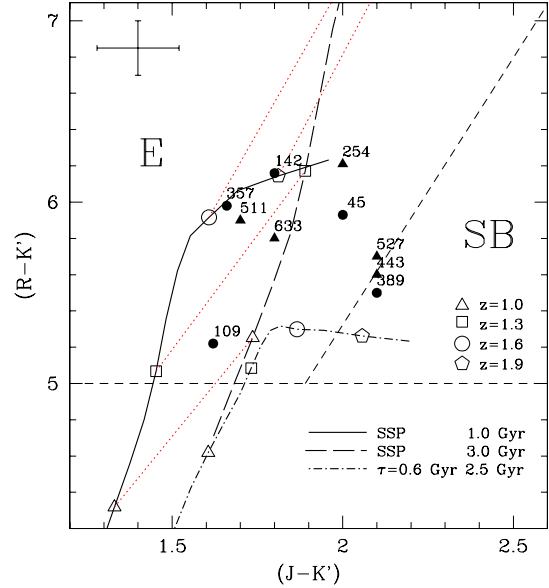}
 \caption{Color-colour diagram for the 10 early-types of our sample of EROs
(triangles stand for lower limits in the R-K' colour).
The dashed thin lines define the Pozzetti and Mannucci (2000) classification
criterion  by separating two regions in the diagram
where ellipticals (E) and starburst galaxies (SB) among the EROs are expected to fall.
The three curves represent the colour-colour tracks for two SSPs
(BC03) 1 and 3 Gyr old respectively,
and for an exponentially declining SFH with time scale
$\tau=0.6$ Gyr and 2.5 Gyr old,
seen at $1<z<2$. Salpeter IMF at solar metallicity has been assumed.
The empty symbols mark the different redshift along the tracks,
and dotted (red) lines connect points with the same redshift
along the SSPs tracks.
Typical measurement errors are shown in the left corner.}
\end{figure}

\section{Comparison with models}

In the following we make use of stellar population synthesis
models to constrain age, K'-band luminosity and stellar
mass content of the 10 early-type galaxies. To this end,
each observed SED (described by the broad band photometry and
by the near-IR spectra) has been compared with two sets of 
spectro-photometric models.
One set has been derived from the latest
version of the BC03 code, assuming
4 different Initial Mass Functions (IMFs, 
0.1M$_\odot<\mathcal{M}<100$M$_\odot$): Salpeter (Sal; Salpeter 1955),
Miller-Scalo (MS; Miller \& Scalo 1979), Scalo (Sca; Scalo 1986) and
Kroupa (Kro; Kroupa 2001); 6 different star formation histories
(SFHs): a Simple Stellar Population (SSP) and 5 SFHs described by exponentially
 declining star formation rates (SFRs) 
with e-folding time $\tau$ in the range 0.1-2 Gyr;
3 different metallicity values. 
A second set of templates is based on the SSP derived by Maraston (2005: Ma05),
assuming Salpeter IMF and solar metallicity.
The basic models parameters are summarized in Table 2.

\begin{table}
\caption{Parameters used to construct the grid of models
adopted in the $\chi^2$-minimization procedure}
\centerline{
\begin{tabular}{ll}
\hline
SFH $\tau$ [Gyr] & SSP, 0.1, 0.3, 0.6, 1, 2 \\
\hline
Metallicity $^{(a)}$ & 0.2 $Z_\odot$, 0.4 $Z_\odot$, $Z_\odot$\\
\hline
Extinction law & Seaton (1979); Calzetti et al. (2000)\\
\hline
IMFs $^{(b)}$ & Sal, Sca, MS, Kro \\
\hline
\end{tabular}
}
{\it Note}: (a) for Ma05 models only $Z_\odot$\par
\hskip 0.8truecm (b) for Ma05 models only Sal
\end{table}

Ma05 code differs from the BC03
one according to the integration method, and it takes into account in a more
reliable way the late phases of stellar evolution such as the Thermally-Pulsating
Asymptotic Giant Branch (TP-AGB), the blue Horizontal Branch and the very hot old stars.
The inclusion of the TP-AGB produces spectra redder than
those derived by the  BC03 code when ages younger than
1 Gyr are considered. However, the differences in the resulting spectral shape
between SSPs with and without TP-AGB become appreciable at $\lambda > 1 \mu$m
(see figure 14 of Ma05). Since the present analysis is based
on spectral data sampling the rest frame at $\lambda < 1 \mu$m, we do not expect 
different results from the two different codes except when we extrapolate
at longer wavelength (see below for further discussion). 
Moreover, also
the peculiar spectral features at $\lambda < 1 \mu$m 
resulting from the inclusion of the TP-AGB phase
could not be revealed from the very low resolution of the present near-IR spectra.

For each set of models and for each possible combination
of models parameters,
the $\chi^2$-minimization procedure of {\em hyperz} (Bolzonella et al. 2000)
has been used to find the spectral templates which best fit the whole
SED of the galaxies. 
In practice, for each single combination of SFH, IMF and metallicity value
defining a model, we find the best fitting 
template, i.e. the corresponding age of the model for the given redshift of the galaxy.
In the best fitting procedure the extinction 
has been allowed to vary
within 3 ranges (A$_{V}$ = 0$\div$0, 0$\div$0.5, 0.5$\div$1.0) and
at each $z$ galaxies have 
been forced to have ages lower than the Hubble time at that $z$.
We repeated this procedure for the selected IMFs, SFHs, metallicity
and dust values considered, thus associating at each galaxy a wide 
set of synthetic templates. Among all these collected
best fitting templates, we further select the acceptable ones
defined as those which provide
fluxes within 1 sigma (observational error) from the observed ones.
When more than one range of dust extinction values gave acceptable
fit, we selected only the one with the lowest value of A$_{V}$. 
This choice is due to the early-type nature of these galaxies,
and to the fact that dust is almost absent in the local sample of early-type
galaxies.
In most of the cases we could accept A$_{V}=0$, while only for
three galaxies (S2F1\_443, S7F5\_254, S7F5\_45)  $0.0 < A_{V} < 0.5$.
In these latter cases, we verified that the final results
do not depend at all from the choice of the extinction law
adopted to describe the dust reddening. This was expected since 
the different extinction curves have about the same shape
over the range 0.2 $\mu$m - 1.0$\mu$m corresponding to 
the rest-frame spectral range used in the present fitting procedure.

On the basis of the selected set of acceptable templates, for each
galaxy we have defined the range of variability of the parameters which can
be derived from the corresponding accepted models: 
K'-band luminosity, stellar mass content and
mean age of their stellar population.
The resulting values  are summarized in Table 3, where the spanned
ranges are reported, together with the spectroscopic measured redshift.
The limited range reported for each quantity
corresponds to all
the models providing acceptable fits to the
spectrophotometric properties of the galaxies, being
the acceptance criterion
based on the accordance within 1 $\sigma$ between the observed
and the fitted spectral fluxes. 
In this sense, the wide set of model parameters which have been adopted to
perform the fit of the galaxies SED assures that the derived ranges
of each measured quantity are robust estimates of their uncertainty.
Apart from this uncertainty, each value of M$_{K'}$ and $\mathcal{M}_{star}$
within the acceptable range has its own statistical error,
due to the errors in the apparent K'-band magnitude ($\la 0.1$),
in the redshift measure ($\delta z=0.05$, that can be transformed in $\sim0.1$ mag
on the absolute magnitude estimate),
in the adopted k-correction and dust extinction factor ($\sim0.1$mag),
in the assumed $\mathcal{M}$/L$_K'$ value (15\% within a single IMF, 40\% if all the IMFs
are considered).
The corresponding average statistical error on the absolute magnitude   
M$_{K'}$ is 0.2 mag, while the uncertainty on the $\mathcal{M}_{star}$
estimate is $\sim20$\% for fixed single IMF and $\sim45$\% when
all the IMFs are considered. No statistical error is affecting the age parameter
since it has been estimated only by means of models comparison without
involving any measured quantity.

The absolute K'-band magnitudes have been calculated assuming the k-corrections 
derived by redshifting each of the accepted best-fitting templates.
In the case of S2F1\_443, S7F5\_254, S7F5\_45, for which the accepted fits
of their SEDs assume A$_{V}>0$, we have taken into account also the
extinction applied to the templates to fit the data. 
We like to note that the small differences with some of 
the K'-band absolute magnitudes reported in Table 1 of Saracco et al. (2005) are
due to the choice of a unique IMF (Salpeter) in that analysis. 
Moreover, the errors reported by Saracco et al. (2005) are those due to 
the apparent K' band magnitude and redshift uncertainties.

Stellar masses $\mathcal{M}_{star}$ have been derived by the K'-band absolute 
magnitudes by means of the mass-to-light ratio $\mathcal{M}$/L$_K'$
relevant to the accepted best-fitting models.
The largest uncertainty in the stellar mass computation comes from the variation 
of $\mathcal{M}$/L according to the age of the stellar population 
and to the adopted IMF. In Table 3, we report the range of values
of $\mathcal{M}_{star}$ obtained with the whole set of IMFs listed in Table2,
with the Salpeter IMF alone and with the Kroupa IMF alone, in columns 5, 6 and 7 respectively.
It appears that on average the stellar masses derived assuming
Salpeter IMF are larger than those obtained with Kroupa IMF of about a factor 1.3.
For comparison, with the Chabrier (2003) IMF, that has been recently widely accepted
as a good universal parametrization of this function (see BC03), we would have
derived stellar masses about a factor 1.5 smaller than those derived with Kroupa IMF
and a factor 2 smaller than those derived with Salpeter IMF (e.g., Bundy et al. 2005).
It can be seen that, in spite of the large
grid of models considered, the stellar mass varies within a narrow
range and typically  within a factor 3 even when some dust
extinction is present.

As to the age of the stellar population of our galaxies,
the age $t$ of the best fitting models cannot be considered its reliable
estimator.
Indeed, the spectro-photometric properties of a stellar population 
are basically defined by the ratio {\it t}$/\tau$, i.e. the ratio
between the age and the SF time scale. 
This implies that the best-fitting age {\it t} exhibits an obvious
degeneracy with respect to $\tau$: similar stellar populations can be  
equivalently described by young ages {\it t} in models with short SF 
time scale (small value of  $\tau$) or by older ages {\it t} in models with 
longer SF time scales. 
Thus, in order to derive a robust estimate of the age of the bulk of 
stars in galaxies,  we derived the mass weighted age
$age_w$ for each best-fitting template defined as:

\begin{equation}
age_w = {{\int_{0}^{t_{temp}}{SFR(t_{temp}-t') \ t'}dt'}
\over 
{\int_{0}^{t_{temp}}{SFR(t')}dt'}}
\end{equation}

\noindent
where $t_{temp}$ is the age  of the template. 
Any template defined by a fixed value of $t_{temp}$  and SFR(t') 
can be seen as the sum of SSPs with different ages.
Each SSP provides a fraction of the total mass which depends on its own
age ($t_{temp}$ -t') and on the SFH describing the template itself.
The mass weighted age is obtained by summing the ages of 
the SSPs, each of them weighted on its mass fraction.
Nevertheless, even the mass weighted age
exhibits a degeneracy, the one with metallicity: young stellar populations
at high metallicity have spectro-photometric properties similar
to those of older stellar populations at lower metallicity.
In Table 3, we report the range of values
of $age_w$ obtained from the best-fitting templates at solar metallicity
(i.e., expected lower limits with respect to the age-metallicity degeneracy) and
at Z$<$Z$_{\odot}$. We like to note that the upper limit of the
accepted range of this parameter is partly due to the
imposed constraint on the fitting procedure, for which models age 
cannot be higher than the Hubble time at each redshift. In some cases,
if this constraint is relaxed, older ages are obtained.
We verified that ages older than the universe, when selected by the fitting 
procedure,  only marginally improve the goodness of the fit obtained
with younger ages. Moreover, similar results can be obtained 
assuming small amount of dust reddening and/or metallicity slightly
higher than the solar value. The available data are not
detailed enough to reliable constrain the metallicity and
the dust content of the galaxies. Thus, 
since these older ages would only strengthen our conclusions on the existence
of old stellar populations in massive galaxies at $z\sim1.5$ which
will be discussed in the next section, 
we prefer to obtain meaningful ages of the galaxies
by imposing the limit of the Hubble time at each redshift
in the fitting procedure.

\begin{table*}
\caption{Main physical parameters of the 10 early-type 
galaxies, derived by comparison with the grid  
of spectro-photometric templates based on BC03 
and M05 models. For each parameter the range of values
covered by the accepted best-fitting templates is reported. 
M$_{K'}$ is given separately for the two sets of models, while $age_w$
is reported for solar and sub-solar matallicity.
In the second column, the spectroscopic redshift is also reported.
Statistical errors are not included and have to be associated to each
value within the reported ranges (see details in the text). }  
\centerline{
\begin{tabular}{lcccccccc}
\hline
  \hline
(1)&(2)&(3)&(4)&(5)&(6)&(7)&(8)&(9) \\
\hline
  Object  & $z_{spec}$ & M$_{K'}$ & M$_{K'}$ & $\mathcal{M}_{star}$ & $\mathcal{M}_{star}$ &  $\mathcal{M}_{star}$ & $age_w$ & $age_w$ \\
          &     & BC03 & Ma05    & [10$^{11}$ M$_{\odot}$] & Sal [10$^{11}$ M$_{\odot}$] & Krou [10$^{11}$ M$_{\odot}$] & Z$_{\odot}$ [Gyr] & Z$<$Z$_{\odot}$ [Gyr]  \\
  \hline
S2F5\_109  & 1.22$\div$0.05 & -27.4$\div$-27.2 & -27.4$\div$-27.2 & 4.7$\div$14.6  & 9.3$\div$12.8  & 7.0$\div$14.6  & 1.4$\div$2.0 & 2.2$\div$3.2 \\
S7F5\_254  & 1.22$\div$0.05 & -26.4$\div$-26.2 & -26.4$\div$-26.2 & 4.5$\div$\ 9.0 & 8.4$\div$\ 9.0 & 6.5$\div$\ 7.2 & 4.9$\div$5.1 & 4.9$\div$5.1 \\
S2F1\_357  & 1.34$\div$0.05 & -26.5$\div$-26.3 & -26.4$\div$-26.2 & 4.2$\div$\ 9.4 & 7.9$\div$\ 9.4 & 6.3$\div$\ 7.2 & 3.9$\div$4.1 & 3.9$\div$4.1 \\
S2F1\_527  & 1.35$\div$0.05 & -26.3$\div$-25.7 & -26.3$\div$-25.7 & 2.4$\div$\ 5.9 & 4.8$\div$\ 5.9 & 3.7$\div$\ 4.6 & 4.4$\div$4.6 & 3.4$\div$4.6 \\
S2F1\_389  & 1.40$\div$0.05 & -26.5$\div$-25.9 & -26.5$\div$-25.9 & 1.7$\div$\ 5.6 & 2.4$\div$\ 5.6 & 1.9$\div$\ 4.4 & 2.5$\div$3.5 & 1.5$\div$3.5 \\
S2F1\_511  & 1.40$\div$0.05 & -26.2$\div$-26.0 & -26.5$\div$-26.3 & 1.1$\div$\ 5.5 & 2.2$\div$\ 5.5 & 1.7$\div$\ 4.3 & 1.0$\div$1.6 & 1.5$\div$4.5 \\
S2F1\_142  & 1.43$\div$0.05 & -26.6$\div$-26.4 & -26.6$\div$-26.4 & 3.1$\div$\ 9.3 & 5.8$\div$\ 9.3 & 4.4$\div$\ 7.2 & 2.0$\div$2.4 & 2.5$\div$3.5 \\
S7F5\_45\  & 1.45$\div$0.05 & -27.0$\div$-26.6 & -27.2$\div$-27.0 & 2.3$\div$\ 8.7 & 4.6$\div$\ 8.7 & 3.4$\div$\ 7.0 & 1.4$\div$2.0 & 1.4$\div$3.0 \\
S2F1\_633  & 1.45$\div$0.05 & -26.3$\div$-26.1 & -26.3$\div$-26.1 & 3.1$\div$\ 7.5 & 5.6$\div$\ 7.5 & 4.2$\div$\ 5.9 & 3.5$\div$4.5 & 3.5$\div$4.5 \\
S2F1\_443  & 1.70$\div$0.05 & -26.8$\div$-26.5 & -26.5$\div$-26.3 & 2.0$\div$\ 9.4 & 3.8$\div$\ 9.4 & 3.1$\div$\ 7.6 & 3.2$\div$3.8 & 1.5$\div$3.5 \\
\hline
\hline
\end{tabular}
}
\end{table*}

\vskip 0.2truecm
As expected, the two sets of models (BC03 and Ma05) lead to similar results in the
$age_w$ and $\mathcal{M}_{star}$ estimate of all the galaxies. Differences can be
appreciated only when the absolute magnitude M$_{K'}$ is derived, since its value
is calculated from the ratio between the flux at $\lambda_{restframe}\sim 0.9\mu$m
(i.e., the apparent K' flux at $z\sim1.5$) and the flux at $\lambda_{restframe}\sim 2.2\mu$m.
As already noted, models Ma05 differs from BC03 at $\lambda > 1.0\mu$m and as a consequence
the {\it k-}corrections in the K' band are different in the two models. 
Differences in the K' band {\it k-}corrections depend from the age of the stellar populations
and  at $z\sim1.5$ can be 0.3-0.4 mag for SSP 1 Gyr old. 
As can be noted from Table 3 (where the M$_{K'}$ range is reported
separately for the two sets of models), the resulting
differences in the derived acceptable range of M$_{K'}$ are smaller. 
This is due to ages higher than 1 Gyr for most of the galaxies and to
slightly different ages of the Ma05 models selected by the best-fitting procedure
(to fit the whole SED) with respect to those selected among the BC03 models. 
We like to emphasis that even when the
derived absolute magnitudes between the two models are different, the 
related stellar mass estimates $\mathcal{M}_{star}$ 
do not exhibit the same difference, 
since the $\mathcal{M}$/L$_K'$ ratio is consequently different.

\vskip 0.2truecm
Finally, we like to recall that in the models selection for each galaxy, we kept
the lowest value of dust reddening needed to obtain acceptable fits.
For some galaxies for which we have excluded A$_{V}>0$ from the accepted
models, fits with some dust reddening were in principle good enough to be selected.
If we would have included also these fits, the acceptable ranges of their luminosities
and stellar mass content would have been enlarged on the side of their upper limit.
Furthermore, possible younger ages could have been included. This would have strengthened 
our conclusions on the existence of very massive galaxies at $z\sim$1.5, even if admitting
a larger range of age values. 
Since the data we have at hand do not allow us to reliable estimate the possible dust 
reddening affecting our galaxies, we prefer to be conservative in accepting the minimum 
value of the extinction requested to fit their SEDs, considering their early-type
nature.

\section{Results and discussion}
In the following sections, we analyze the main properties of the
10 early type galaxies summarized in Table 3.
In particular, we discuss the possibility
of a common origin and/or evolution (\S4.1) and
we compare their properties with those of local galaxies (\S4.2).

\subsection{Early-type galaxies at $z\simeq1.5$: evidences for different 
SF histories?}
The bulk of stars in 6 out of the 10 galaxies is $\sim$3-5 Gyr old
while the other 4 show mean stellar ages of about 1-2 Gyr, even assuming
the highest value of metallicity adopted in our fitting procedure.
Figure 3 shows this evidence taking also
into account the different redshift of the galaxies (i.e., different
age of the Universe at their redshift).
For each galaxy we computed the quantity

\begin{equation}
\Delta Age = Age_{Universe}(z_{gal}) - Age_{gal}
\end{equation}

\noindent
defined as the difference between the age of the Universe at the redshift
of the galaxy and the age of the stellar population of the galaxy itself
as derived in \S 3, and the corresponding formation redshift $z_{f}$:

\begin{equation}
z_{f}=z[Age_{Universe}(z_{gal}) - Age_{gal}]
\end{equation}

\noindent
In the left panel of Figure 3, the quantities defined above are
plotted against the redshift of the galaxies.
Values of $\Delta$Age close to 0 indicate that stars are as old
as the Universe at the galaxy redshift, i.e., their formation
redshift $z_{f}$ is high. 
Larger values of $\Delta$Age means that the
stellar populations are, on average, younger than the Universe at
the galaxy redshift 
and that at least a fraction of them formed recently.
As to our galaxies, the figure clearly shows a quite large
spread in their age and formation redshift. Indeed, 
the stellar content of 3 of them (S2F5\_109, S2F1\_511, S7F5\_45)
appears to be formed when the Universe was about 2.5 - 3.0 Gyr
old (i.e., $z_f \approx\ $ 2), 2 galaxies (S2F1\_142 and S2F1\_389)
show the bulk of their stellar
content as formed when the Universe was about 2 Gyr old (i.e., $z_f \approx\ $ 3)
while the remaining 5 (S7F5\_254, S2F1\_633, S2F1\_357, S2F1\_443 and S2F1\_527)
present a stellar content not younger than 1 Gyr with respect to the Universe
(i.e., as formed at $z_f >$ 5).
The right panel of the Figure 3 represents these results
by showing the age of the Universe at $z_{f}$ (squares) and at $z$ (triangles)
for each of the galaxies. The length of the line connecting the two points
corresponds to the age of their stellar population, while the upper scale
helps in relating the age of the Universe at $z_{f}$ and $z_{f}$ itself. 
Again it can be noticed that all the galaxies have $z_{f}>2$, and that for
the oldest ones (i.e., those represented by the longest lines, namely
S7F5\_254, S2F1\_633, S2F1\_357, S2F1\_443 and S2F1\_527) the formation redshift
is $>5$. These different formation redshift derived for the galaxies
of the sample suggests
that either {\it i}) the bulk of stars in the 
various galaxies formed at different epochs, i.e. different 
early-type galaxies experienced 
their main burst at different redshift, or {\it ii})  some of them have experienced 
later bursts of star formation after the main event, i.e., they
are characterized by longer SF time scale, and
the derived age parameter is not corresponding to the true value.
In the first hypothesis, the main star formation event
from which the stellar content of massive early-type galaxies originated
took place at different
epochs for different galaxies at $z_f>2$. 
The second hypothesis is based on the fact
that most of the spectro-photometric data we used to constrain
the stellar ages of our galaxies sample the rest frame UV and optical
range, which are very sensitive to the presence of even small
amounts of  young ($<$ 1 Gyr) stars.
In other words, even a small fraction  of stars younger than 1 Gyr
may contribute for more than 90\% to the luminosity at
$\lambda < 0.3 \mu m$ and up to 50\% at $0.3 \mu m < \lambda < 0.5 \mu m$,
mimicking a young age for the whole population.
This can be seen in  Figure 4 where we show
the synthetic spectrum of a composite stellar population (thick black line)
resulting from the sum of a 3 Gyr old SSP providing 90\% of the 
stellar mass (thin black line) and of a
0.5 Gyr SSP accounting for 10\% of the stellar mass (green/grey line).
\begin{figure*}
 \centering
 \includegraphics[width=16.5cm]{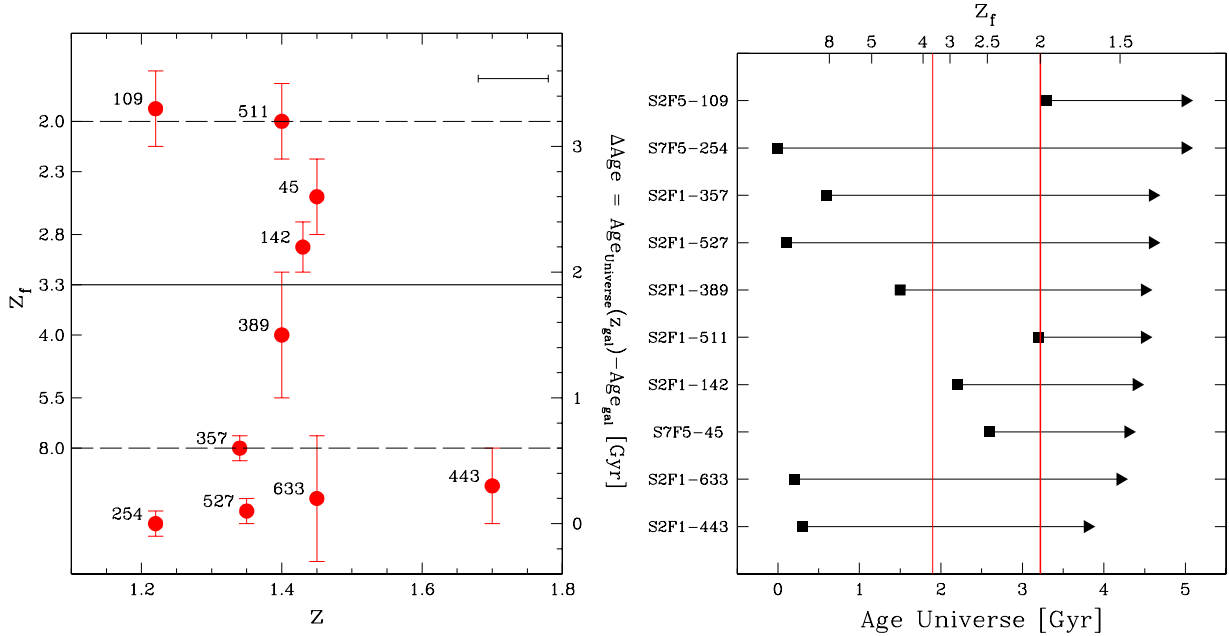}
 \caption{Left panel: for each galaxy the redshift of formation 
($z_{f}$) of its stellar population is shown as a function of the redshift itself.
Label and scale on the right side of the figure give the corresponding 
value of the difference between the age of the Universe
at the galaxies redshift and the age of the bulk of their stellar content.
The value
of 1.9 $\pm 1.3$ Gyr (thick black lines) can be regarded as the age of
the Universe at the time of the strong SF episode from
which the present sample of early-type galaxies formed, and it corresponds
to $z_{f}\sim3.3$.
Right panel: for each galaxy the age of the Universe at its redshift
(triangles) and its formation redshift $z_{f}$ (squares) as deduced
from the mean age of its stellar content are reported.
The length of the line connecting each points pair represents
the mean age of the stellar populations of the 10 galaxies.
Labels on the left y-axis identify the galaxies. Label and
scale on the bottom x-axis refer to the age of the Universe 
in Gyr, while on the top they refer to the formation redshift.
The formation redshift of all the galaxies results $>2$ (indicated
by the thick red/grey vertical line)
with an average value $z_{f}\sim3.3$ corresponding to a 1.9 Gyr old Universe
(thin red/grey vertical line). All the 6 galaxies with $z_{f}> 3.3$ have
stellar populations older than 3 Gyr, while the remaining 4 have
mean ages of about 1.5 Gyr.}
\end{figure*}
\noindent
It is clear from the figure that the young population dominates or becomes
comparable to the old component contribution in the emission
at $\lambda<0.5$ $\mu$m in spite of the negligible  mass fraction.
This shows that while we can be confident that old age estimates
(i.e., $>$ 3 Gyr, $z_f\ge3$) most probably approaches the real age
of the stellar bulk of the galaxies, young values could be those
characterizing only a small fraction of their stellar content.
Thus, some  of the early-types can have experienced secondary
weak star formation events during their evolution.
In particular, S2F5\_109, S2F5\_45, S2F1\_511
for which we derived $age_w < 2$Gyr and $z_f<3$ could just suffered
minor star forming events at that redshift while the bulk of their
stellar content could be formed at higher redshift.
The data we have in hand do not allow us to discriminate
between true young ages of the stellar content of galaxies and
recent minor burst superimposed on a much more massive content of
old stars,
i.e. we are not 
able to distinguish between different epochs of formation and/or 
different star formation histories. 
However, these results indicate that the evolution of 
the population of $z \sim 1.5$ field massive early-type galaxies 
is not unique but characterized by  different SFHs, and that 
the formation  could be  distributed in a range of redshift at $z>2$.
\begin{figure}
\centering
\includegraphics[width=9.5cm]{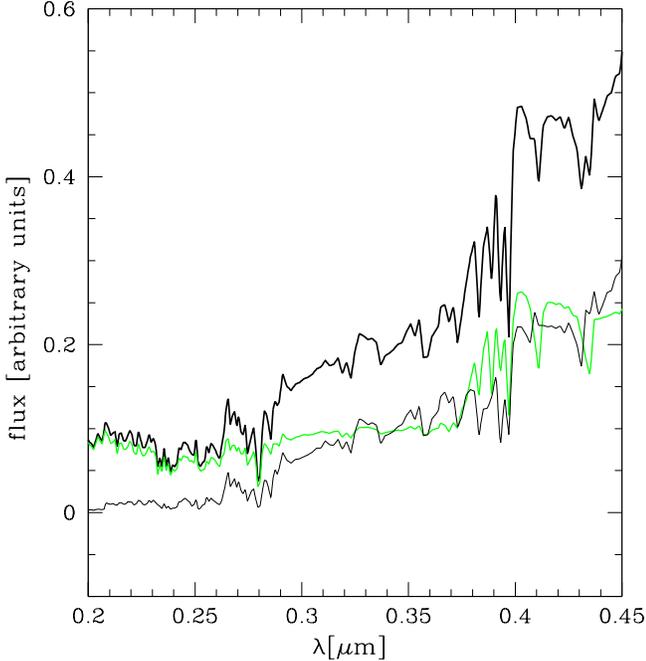}
\caption{Thick black line: synthetic spectrum of a composite
stellar population obtained by summing 90\% of the mass
of a 3 Gyr old SSP (thin black line) and 10\% of the mass 
of a 0.5 Gyr old SSP (green/grey line).
Both the SSPs correspond to solar metallicity and to
the Salpeter IMF.  The contribution to the total
luminosity of the young SSP is more than 90\% at
$\lambda < 0.3 \mu m$, and similar to the contribution
of the older much more massive component at
$0.3\mu m < \lambda < 0.5 \mu m$.
}
\end{figure}

\subsection{$\Delta$(4000\AA): comparison with local early-type galaxies}
Given the rest-frame absolute  magnitudes of the 10 galaxies 
(M$_K'\le-26.0$) and assuming that they evolve passively from $z\simeq1.2$ 
(the minimum redshift of our sample)
to $z=0$ ($\Delta K'\simeq1.2$ mag), the resulting K'-band luminosity of these 
early-type galaxies at $z=0$ would be L$_{z=0}\ge2$L$^*$.
Thus, these galaxies are the high-z counterpart of the most massive 
early-types populating the bright end of the 
local luminosity function of galaxies. A comparison
between the properties of this class of galaxies
at $z\sim1.5$ and at $z=0$ provides constraints on their
evolution.

\vskip 0.4truecm
\noindent {\it Stellar Age} -
Since the local Universe is $\sim8-9$ Gyr older than the Universe at  
$z\sim1.5$, we expect to observe differences between the properties of our
early-type galaxies and those at $z\sim0$ consistent with this difference
in their age.
Actually, the stellar population of local high mass early-type galaxies
seems to have ages of about 10 Gyr or more (e.g. Caldwell et al. 2003).
These values, derived by means of detailed and accurate analysis
of the galaxies spectral narrow-band indices, can be
directly compared with those of $age_w$ derived for our sample
of galaxies. Indeed, at ages greater than 10 Gyr differences between the
model parameter age and the quantity $age_w$ defined above are less than
1 Gyr, even for long SF time scale (e.g., $\tau=1$Gyr).
Our galaxies at $z\sim1.5$ are $1\div 5$ Gyr old, in agreement
with the younger age of the Universe at that redshift.
Thus, the SF history followed on average by these massive 
galaxies has to be consistent
with the observed nearly passive aging.

\vskip 0.4truecm
\noindent {\it $\Delta$(4000\AA)} -
The available spectral data of the 10  galaxies 
can be used to construct the average (rest-frame) optical spectrum
of early-type galaxies at the mean redshift $z=1.5$,
that can be compared with the SED of
their local counterparts. In the previous section (\S4.1) we have
discussed the spread in age of the stellar content of
the galaxies in the sample. Since different stellar ages
mean different spectro-photometric properties, we have averaged
the spectra of the 10 galaxies after dividing them according to their ages:
{\it i) (old)} composed by galaxies with {\it age$_w$} $\ge 3.0$ Gyr 
(namely S7F5\_254, S2F1\_527, S2F1\_389, S2F1\_633, S2F1\_443, S2F1\_357) and
{\it ii) (young)} composed by galaxies with {\it age$_w$} $< 3.0$ Gyr
(namely S2F5\_109, S2F1\_511, S2F1\_142, S7F5\_45). 
Since among the six {\it old} galaxies only two (S2F1\_633 and S2F1\_443) are
at $z\ga1.4$,
the average spectrum derived by this group of galaxies starts
at $\lambda_{restframe} > 0.40 \mu$m, while that of the {\it young}
one at $\lambda_{restframe} > 0.38 \mu$m.
In Figure 5 we show the two average spectra reported at $z=0$
normalized around $\lambda\sim0.5\mu$m rest-frame: thick black line stands for
the {\it old} spectrum  while thin black line for the {\it young} one.
For comparison we also report (thick green/grey line) 
the mean observed spectrum of local 
elliptical/S0 galaxies by Mannucci et al. (2001, Man01).
In the top right corner, besides the  Man01 spectrum (thick green/grey line), 
we show the synthetic spectra of SSPs 2 Gyr old (thin black line) 
and 4 Gyr old (thick black line),
derived from the BC03 models, adopting solar metallicity
and Salpeter IMF.

\begin{figure}
\centering
\includegraphics[width=9.5cm]{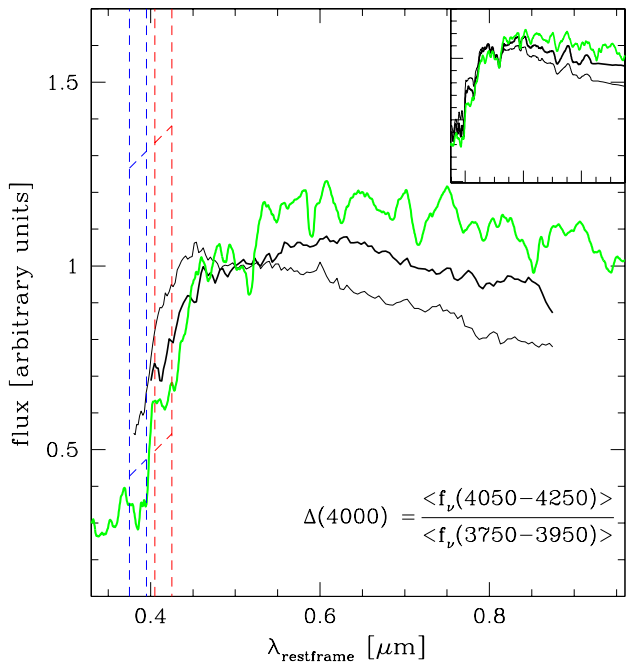}
\caption{Average spectrum of the galaxies
with {\it age$_w$} $\ge 3.0$ Gyr (thick black line, {\it old}) and 
with {\it age$_w$} $< 3.0$ Gyr (thin black line, {\it young}).
The {\it old} spectrum starts
at $\lambda > 0.4 \mu$m because 4 out of the 6 galaxies
of this group are at $z\la1.4$.  
For comparison, the mean observed spectrum of local early-type
galaxies (elliptical and S0)
by Mannucci et al. (2001, Man01) is also shown (thick green/grey line).
All the spectra are normalized around 0.5$\mu$m rest-frame.
In figure the adopted definition of the $\Delta$(4000\AA) index
is reported (Bruzual 1983) together with the relevant spectral bands
(dashed lines). In the top right corner, synthetic spectra
of two SSPs 4 Gyr old (thick black line) and 2 Gyr old (thin black line)
are shown on the same spectral range together with the Man01 spectrum of
local early-types.
}
\end{figure}

The most striking feature, apart from the overall similarity,
is the steeper J-K' slope shown by both the two composite spectra
of early-type galaxies at $\sim 1.5$ with respect to that at
$z=0$. For the {\it young} spectrum it is also evident a
smaller 4000\AA\ break with respect to the Man01 spectrum. Indeed,
the flux of the spectrum at $z\sim1.5$ below 4000\AA\  (rest-frame)
is about twice that of the local one, while above
4000\AA\ the ratio between the flux of the two spectra is less than 1.5.
Both these two features are clear signs
of younger ages at $\sim 1.5$ with respect to $z=0$,
as can be deduced looking at the synthetic spectra in the top
right corner of the figure where the 2 Gyr SSP and 4 Gyr SSP
well reproduce the overall continua of the {\it young} and
{\it old} spectra respectively. 
In order to better quantify this evidence, we have estimated the 
$\Delta$(4000\AA) index of the average spectra at $z=0$ and
at $z=1.5$ (from the {\it young} spectrum only, since the {\it old}
one does not cover the needed blue part of the spectral range due
to its slightly lower mean redshift) 
adopting the definition of Bruzual (1983).
While massive early-type galaxies at $z\sim1.5$ are characterized by 
$\Delta$(4000\AA)$_{z\sim1.5}$=1.7,
their local counterparts have an index $\Delta$(4000\AA)$_{z=0}$=2.2.
Thus, the SFH followed by massive field early-type galaxies
has to be able to produce during an interval of about 8-9 Gyr
(i.e., from $z\sim1.5$ to $z=0$)
the observed variation in the $\Delta$(4000\AA) index.

Figure 6 shows the expected values of the $\Delta$(4000\AA) index
as a function of  the mass weighted age $age_w$ of the stellar populations
(scale and label on the bottom). 
For useful comparison, scale and labels on the top
give the correspondence between $age_w$ and $z$ assuming $z_f=4$.
Our sample of early-type galaxies is at $z\sim1.5$ with a typical
mean age of $log(age_w)=9.4$.
In the upper panel, we show the index value as a function
of $age_w$ for models
with different IMFs while the SFH is described by an 
exponentially declining SFR with time scale $\tau=1$ Gyr
(solar metallicity).
It is clear that the $\Delta$(4000\AA) index is in fact independent of the IMF.
For this reason, the models shown in the other two panels have been obtained  
considering the Salpeter IMF only.
In the middle panel the dependence of the index on the metallicity
for  fixed SFH ($\tau=1$ Gyr)
is shown, while the lower panel presents
the dependence on different SFHs for fixed metallicity (Z=Z$_\odot$). 

\begin{figure}
\centering
\includegraphics[width=9.5cm]{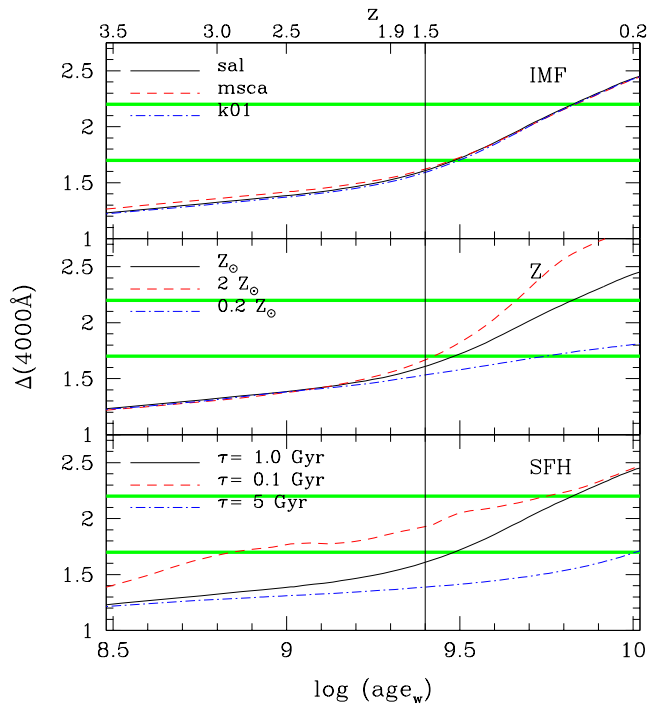}
\caption{{\em Upper panel}:
4000\AA\ break as a function of $age_{w}$ for different IMF
and SFH described by $\tau = 1$ Gyr 
(solar metallicity). 
{\em Middle panel}: 4000\AA\ break as a function of $age_{w}$ for different metallicities
and SFH described by $\tau = 1$ Gyr (Salpeter IMF).
{\em Lower panel}: 4000\AA\ break as a function of $age_{w}$ for different SFHs
(Salpeter IMF and solar metallicity). {\it Note}: on the top of the
figure, label and scale give the $z$ values corresponding to the age ones, 
assuming a formation redshift of the stellar population $z_{f}=4$
(i.e., 1.5 Gyr younger than the Universe at the same $z$). The
early-type galaxies analyzed in the present work are at $z\sim1.5$ and
about 3 Gyr old on average (i.e., $log(age_{w})=9.4$, vertical line).
Green/grey thick horizontal lines mark the $\Delta$(4000\AA) index values
at $z\sim1.5$ ($\Delta$(4000\AA)=1.7) and at $z=0$ ($\Delta$(4000\AA)=2.2).}
\end{figure}

It can be seen that metallicity Z$\ll$Z$_\odot$ produces a 
$\Delta$(4000\AA)  index much smaller than those observed both at high and 
at low redshift and fails to produce  the observed variation (about a factor 
$1.3$) from $z\sim1.5$ to $z=0$.
Analogous arguments apply also to metallicity Z$\gg$Z$_\odot$, while
Z$\la$Z$_\odot$ better matches the observations.
As to the SFH, for the same value of $age_w$  an almost coeval stellar 
population ($\tau$=0.1Gyr) is always characterized by a $\Delta$(4000\AA) 
much higher than that of populations  with larger age spread 
($\tau>$1.0 Gyr).
In particular, SFHs described by times scales $\tau>1$ Gyr  fail to
reproduce the observed local value ($\Delta$(4000\AA)=2.2) even for ages 
as old as the Universe (i.e., 13 Gyr).
On the other hand, extremely  short SF time scales ($\tau=$ 0.1 Gyr) would need 
more than 8-9 Gyr (i.e., $>$10 Gyr) to
reproduce the observed variation of $\Delta$(4000\AA) from  
$z\sim1.5$ to $z=0$.
Thus, the more reliable SFHs able to produce the $\Delta$(4000\AA) 
index observed in massive early-types at $z\sim1.5$ and at $z=0$ are those 
characterized by time scales  $< 1$ Gyr
and metallicity  Z$\la$Z$_\odot$.
It is worth noting that these models lead to date the stellar content
of the $z \sim 1.5$ mean early-type galaxy as 2.5 Gyr old, that is consistent with 
the mean age of the single galaxies as derived through the comparison between their individual
spectro-photometric properties and synthetic models of Sec. 3.

\vskip 0.3truecm
On the basis of the mean age of the stellar populations derived
from our analysis (\S 4.1) and on the likely SFHs they could follow (\S 4.2), we can 
constrain the formation of our massive early-type galaxies in the redshift range 
$2<z_f<3.5-4$. This result agrees with the findings of Thomas et al. (2005)
that is based on what they call ``archaeology approach''. Using the models of
absorption line indices based on the Ma05 models, 
they performed a detailed analysis
of spectral properties of 124 local early-type galaxies in different environments,
and they concluded that massive field early-types should have formed
their stellar content around $z\approx2$ with short SF timescales.
Furthermore, analysis of the Fundamental Plane properties of early-type
galaxies at $z\la 1$ in the field (van der Wel et al. 2004; van Dokkum et al. 2003) 
and in clusters (e.g. Bender et al. 1996) show strong evidences of
the formation of their stars at $z\ga2$.
Our results confirm the prediction of the previous works extending 
the analysis of the age and star formation history of field 
early-type galaxies to higher redshift ($1.2 < z < 1.7$).
Indeed, previous studies of field early-type galaxies were
focused on either lower redshift regimes (i.e., $z\sim1$: Mignoli et al. 2005;
Treu et al. 2005; van der Wel et al. 2004; van der Wel et al. 2005) or on
less massive objects (i.e., L$<$L$^*$: Daddi et al. 2005; McCarthy et al. 2004).
We like to note that the above results imply that old ($>$1 Gyr)  
massive ($\mathcal{M}_{star}>10^{11}$ 
M$_\odot$) early-type galaxies should be observed fully assembled at redshift 
$z\ga 2$ as well as the progenitors  characterized by very high 
star formation rates.
In fact, some observational evidences seem to confirm these expectations.
Early-type galaxies with stellar masses of the order of 
$10^{11}$ M$_\odot$ and ages of about $\sim$1 Gyr  are being found
at $z\sim2.5-3$ from deep near-IR selected samples (Longhetti et al. 2004; 
Saracco et al. 2004; van Dokkum et al. 2004; Cimatti et al. 2004) 
and some of the SCUBA sources at $z\sim3$
turned out to be massive galaxies with star formation rates of many
hundreds M$_\odot$ yr$^{-1}$ (Genzel et al. 2003; Smail et al. 2002).

\section{Summary and conclusions}
We presented the analysis of 10 massive early-type galaxies 
revealed in a complete sample of 36 bright (K' $<$ 18.5)
EROs (R-K'$>$ 5) selected from the MUNICS survey. 
The low resolution near-IR spectra obtained as part
of the on-going spectroscopic follow-up of the whole sample
identify them as early-type galaxies at redshift 
$1.2<z<1.7$. 
Given their extremely bright K'-band absolute magnitudes, 
their resulting stellar masses are well in excess of $10^{11}$ M$_\odot$
leaving aside any model assumption.

\vskip 0.2 truecm
We compared the broad-band photometry and the near-IR 
spectra with a grid of spectrophotometric models.
By means of a $\chi^2$-minimization procedure, we defined
a limited set of all the synthetic templates which well 
reproduce the spectro-photometric properties of the galaxies.
This set provided us with the acceptable range of possible values 
of their main physical parameters, i.e. K'-band luminosity, 
stellar mass and age of the bulk of their stellar content.
This part of the analysis led to the following results:

\noindent $\bullet$
field massive early-type galaxies at $z\sim1.5$ exhibit an apparent spread 
in the age of their stellar populations. In particular, 
the bulk of stars in 6 out of the 10 galaxies is $\sim$3-5 Gyr old
while the remaining 4 ones show mean stellar ages of about 1.5 Gyr. 

\noindent $\bullet$
the observed spread in age can be explained either assuming a corresponding
spread in the formation redshift or as due to the underestimate
of the real stellar age for the apparent young galaxies.
In the former hypothesis, the formation redshift
of the oldest 6 galaxies is $z_{f}\ga4$ while for the youngest ones is
$2<z_{f}<4$. In the second hypothesis, even the apparent young galaxies
have a stellar content as old as the other galaxies, but it is hidden
by a negligible fraction of young stars formed in a recent ($\sim 1$ Gyr)
star forming episode.
If this is the case, it would be difficult to disentangle the two
populations.
Anyway, these results indicate that the star formation history  of the 
population of field massive early-type galaxies at $z\sim1.5$ is not unique, 
being not quiescent for some of them.

\vskip 0.2 truecm
Furthermore, we averaged
the near-IR spectra of the 10 early-type galaxies after dividing them into two groups: 
{\it i)} the {\it (old)} one composed of galaxies with {\it age$_w$} $\ge 3.0$ Gyr  
and {\it ii)} the {\it (young)} one composed of galaxies with {\it age$_w$} $< 3.0$ Gyr.
Their comparison with the mean observed spectrum of the local early-type galaxies
allowed us to derive the following results:

\noindent $\bullet$
both the two composite spectra of early-type galaxies at $\sim 1.5$ 
show a $\lambda_{restframe}>0.5 \mu$m steeper slope 
with respect to their local counterpart.
The {\it young} spectrum allows also the measure
of the 4000\AA\ break, and its value ($\Delta$(4000\AA)$_{z\sim1.5}$=1.7)
is lower with respect to that
measured on the local template ($\Delta$(4000\AA)$_{z=0}$=2.2).
Both these two features are clear signs
of younger ages at $z\sim 1.5$ with respect to $z=0$,
independently of any model assumption used
to describe the index behavior.

\noindent  $\bullet$
among the SFHs adopted to model the spectral properties
of the galaxies, the more reliable one, capable to reproduce the values 
of the $\Delta$(4000\AA) index both at $z\sim1.5$ and at $z=0$,
is characterized by a SF time scale shorter than 1 Gyr with metallicity 
not higher than the solar value. 

\noindent  $\bullet$
on the basis of the average spectral shape and of the $\Delta$(4000\AA) index
value, the stellar content of the $z\sim1.5$ early-type galaxies is dated as
3 Gyr old and their formation is well described assuming  $2<z_{f}<3.5-4$.
This result is consistent with the stellar ages derived 
in the first part of our analysis based on the study of the individual
spectral continua.

\vskip 0.2 truecm
Our results confirm the predictions of previous works
based on  samples of local field and cluster galaxies, extending
the analysis of the age and star formation history of field
massive early-type galaxies to higher redshift ($1.2 < z < 1.7$).
All these findings imply that old ($\sim$1 Gyr) massive 
($\mathcal{M}_{star}>10^{11}$  M$_\odot$)
early-types galaxies are expected at redshift $z>2$, together with 
their progenitors characterized by high star formation rates.
Recent observational evidences seem to confirm these expectations
suggesting that massive early-type galaxies
formed their stars at $z\sim3-4$ over a time scale of about 1 Gyr or less,
and implying the nearly passive evolution observed from $z\sim1.5$ to $z=0.0$
(i.e., favouring the monolithic scenario).
These results should then be considered as strong constraints
for the models of galaxy formation, and
in particular for the semianalytic codes which attempt the link 
between the baryonic mass assembly and the resulting luminous properties
of the formed galaxies.

\section*{Acknowledgments}
We thank the anonymous referee for useful comments which helped
to improve the paper. 
We thank the staff of the TNG for their support during the
observations.
PS acknowledges a research fellowship from the {\it Istituto Nazionale 
di Astrofisica} (INAF). This work has received partial financial support
from the Italian Ministry of University and the Scientific and
Technological Research (MIUR) through grant Cofin-03-02-23.
The MUNICS project is supported by the Deutsche
Forschungsgemeinschaft, \textit{Sonderforschungsbereich 375,
Astroteilchenphysik}

\label{lastpage}

\end{document}